\documentclass[iop]{emulateapj}

\usepackage{amssymb}

\def\ts     {\thinspace}
\def\kms    {\ts km\ts s$^{-1}$}
\def\etal   {{\rm et\ts al.}}
\def\msol   {$M_{\odot}$}
\def\lsol   {$L_{\odot}$}

\slugcomment{draft version \today, accepted for publication in the Astrophysical Journal}

\shorttitle{PAH and mid-IR continuum emission in GN20}
\shortauthors{Riechers et al.}

\begin{document}

\title{
Polycyclic Aromatic Hydrocarbon and Mid-Infrared Continuum Emission \\ in a $z$$>$4 Submillimeter Galaxy}

\author{Dominik A. Riechers\altaffilmark{1,2}, Alexandra Pope\altaffilmark{3}, Emanuele Daddi\altaffilmark{4}, Lee Armus\altaffilmark{5}, \\ Christopher L.\ Carilli\altaffilmark{6}, Fabian Walter\altaffilmark{7}, Jacqueline Hodge\altaffilmark{7}, Ranga-Ram Chary\altaffilmark{5}, \\ Glenn E.\ Morrison\altaffilmark{8,9}, Mark Dickinson\altaffilmark{10}, Helmut Dannerbauer\altaffilmark{11}, and David Elbaz\altaffilmark{4}}

\altaffiltext{1}{Department of Astronomy, Cornell University, 220 Space Sciences Building, Ithaca, NY 14853, USA; dr@astro.cornell.edu}

\altaffiltext{2}{Astronomy Department, California Institute of
  Technology, MC 249-17, 1200 East California Boulevard, Pasadena, CA
  91125, USA}

\altaffiltext{3}{Department of Astronomy, University of Massachusetts, Amherst, MA 01003, USA}

\altaffiltext{4}{Laboratoire AIM, CEA/DSM-CNRS-Universit\'e Paris Diderot, Irfu/Service d'Astrophysique, CEA Saclay, Orme des Merisiers, F-91191 Gif-sur-Yvette Cedex, France}

\altaffiltext{5}{Spitzer Science Center, California Institute of Technology, MC 220-6, 1200 East California Boulevard, Pasadena, CA 91125, USA}

\altaffiltext{6}{National Radio Astronomy Observatory, PO Box O, Socorro, NM 87801, USA}

\altaffiltext{7}{Max-Planck-Institut f\"ur Astronomie, K\"onigstuhl 17, D-69117 Heidelberg, Germany}

\altaffiltext{8}{Canada-France-Hawaii Telescope, 65-1238 Mamalahoa Hwy, Kamuela, Hawaii 96743-8432, USA}

\altaffiltext{9}{Institute for Astronomy, 2680 Woodlawn Drive, Honolulu, Hawaii 96822-1839, USA}

\altaffiltext{10}{National Optical Astronomy Observatory, 950 North Cherry Avenue, Tucson, AZ 85719, USA}

\altaffiltext{11}{Institut f\"ur Astrophysik, Universit\"at Wien, T\"urkenschanzstra\ss e 17, A-1180 Wien, Austria}

\begin{abstract}

We report the detection of 6.2\,$\mu$m polycyclic aromatic hydrocarbon
(PAH) and rest-frame 4--7\,$\mu$m continuum emission in the $z$=4.055
submillimeter galaxy GN20, using the Infrared Spectrograph (IRS) on
board the {\em Spitzer Space Telescope}. This represents the first
detection of PAH emission at $z$$>$4. The strength of the PAH emission
feature is consistent with a very high star formation rate of
$\sim$1600\,\msol\,yr$^{-1}$. We find that this intense starburst
powers at least $\sim$1/3 of the faint underlying 6\,$\mu$m continuum
emission, with an additional, significant (and perhaps dominant)
contribution due to a power-law-like hot dust source, which we
interpret to likely be a faint, dust-obscured active galactic nucleus
(AGN). The inferred 6\,$\mu$m AGN continuum luminosity is consistent
with a sensitive upper limit on the hard X-ray emission as measured by
the {\em Chandra X-Ray Observatory} if the previously undetected AGN
is Compton-thick. This is in agreement with the finding at
optical/infrared wavelengths that the galaxy and its nucleus are
heavily dust-obscured.  Despite the strong power-law component
enhancing the mid-infrared continuum emission, the intense starburst
associated with the photon-dominated regions that give rise to the PAH
emission appears to dominate the {\em total} energy output in the
infrared.  GN20 is one of the most luminous starburst galaxies known
at any redshift, embedded in a rich protocluster of star-forming
galaxies. This investigation provides an improved understanding of the
energy sources that power such exceptional systems, which represent
the extreme end of massive galaxy formation at early cosmic times.

\end{abstract}

\keywords{galaxies: active --- galaxies: starburst --- 
galaxies: formation --- galaxies: high-redshift --- cosmology: observations 
--- infrared: galaxies}

\section{Introduction}

Submillimeter galaxies (SMGs; see review by Blain et
al.\ \citeyear{bla02}) are thought to be the progenitors of the most
massive present-day galaxies, making them a key ingredient to studies
of massive galaxy formation and evolution through cosmic time. Many
SMGs are outliers on the stellar mass-star formation rate relation at
$z$$\sim$2 (e.g., Daddi \etal\ \citeyear{dad07}), and they typically
are few kpc diameter, intense ($>$500--1000\,\msol\,yr$^{-1}$)
starbursts embedded in $\gtrsim$10\,kpc gas reservoirs (e.g., Riechers
\etal\ \citeyear{rie11a}, \citeyear{rie11b}; Ivison
\etal\ \citeyear{ivi11}). They frequently feature rapid gas
consumption through high star formation efficiencies that are
associated with ongoing major mergers (e.g., Tacconi et
al.\ \citeyear{tac08}). The high abundance of passive, massive
galaxies already by $z$$\sim$2 (e.g., Renzini \citeyear{ren06})
requires a substantial population of $z$$>$3--4 galaxies with vigorous
starbursts. Studies of the so-called high redshift tail of SMGs at
$z$$>$4 thus promise to provide valuable insights toward understanding
the distribution of formation redshifts of massive early-type
galaxies. The few $z$$>$4 SMGs securely identified to date (e.g.,
Daddi \etal\ \citeyear{dad09} [hereafter:\ D09], \citeyear{dad09b};
Capak \etal\ \citeyear{cap08}, \citeyear{cap11}; Riechers
\etal\ \citeyear{rie10}, \citeyear{rie13}) may already be sufficient
to account for known populations of old massive galaxies at $z$=2--3
(e.g., Coppin \etal\ \citeyear{cop09}).  Also, $z$$>$4 SMGs may have a
significant contribution to the contemporaneous comoving star
formation rate density (e.g., D09).

GN20 ($z$=4.055) is one of the intrinsically brightest SMGs known
($S_{\rm 850\mu m}$=20.3\,mJy; Pope \etal\ \citeyear{pop05},
\citeyear{pop06}), and one of only a few spectroscopically confirmed
unlensed SMGs at $z$$>$4. Located in a massive galaxy protocluster
structure ($\sim$10$^{14}$\,\msol; D09; Hodge
\etal\ \citeyear{hod13}), this remarkable galaxy hosts a massive
$M_{\rm H_2}$=1.8$\times$10$^{11}$\,($\alpha_{\rm
  CO}$/1.1)\,\msol,\footnote{$\alpha_{\rm CO}$ is the conversion
  factor from CO luminosity to molecular gas mass (H12).}
14$\pm$4\,kpc diameter molecular gas disk, which accounts for
$\sim$1/3 of the total (dynamical) mass of the galaxy (Hodge
\etal\ \citeyear{hod12} [H12]). This gas reservoir feeds an intense
starburst that is almost entirely obscured by dust at optical
wavelengths within the central $\gtrsim$10\,kpc (e.g., Iono
\etal\ \citeyear{ion06}; Carilli \etal\ \citeyear{car10} [C10]).  This
intense, $>$1000\,\msol\,yr$^{-1}$ starburst will be sufficient to
turn GN20 into one of the most massive galaxies known within
$\ll$1\,Gyr. Due to the high level of obscuration, it is currently not
known if GN20 hosts an AGN, and thus, if the infrared luminosity
($L_{\rm IR}$) used to estimate the star formation rate is entirely
powered by dust-reprocessed light from young stars.

At low and intermediate redshifts ($z$$\sim$2), spectral decomposition
of mid-infrared spectra has proven to be a powerful tool to
investigate the relative contributions of AGN and star formation to
the infrared light in star-forming galaxies (e.g., Armus
\etal\ \citeyear{arm07}; Farrah \etal\ \citeyear{far07}; Pope
\etal\ \citeyear{pop08} [P08]; Veilleux \etal\ \citeyear{vei09};
Petric \etal\ \citeyear{pet11}), but no constraints exist for galaxies
at $z$$>$4 so far.

We here report the detection of 6.2\,$\mu$m PAH emission and
4--7\,$\mu$m continuum emission toward the $z$=4.055 SMG GN20. This is
the highest $z$ detection of PAH emission reported to date, reaching
back to only $\sim$1.5\,Gyr after the Big Bang. This enables us to
better constrain the intensity of the starburst in this galaxy, and to
determine the presence and energy contribution of an AGN in this
intriguing system. We use a concordance, flat $\Lambda$CDM cosmology
throughout, with $H_0$=71\,\kms\,Mpc$^{-1}$, $\Omega_{\rm M}$=0.27,
and $\Omega_{\Lambda}$=0.73 (Spergel \etal\ \citeyear{spe03},
\citeyear{spe07}).

\section{Observations}

Observations toward GN20 ($z$=4.055) were carried out with the
Infrared Spectrograph (IRS; Houck et al.\ \citeyear{hou04}) on board
the {\em Spitzer Space Telescope} on 2009 May 2--4 with a total
observing time of 24\,hr, split up into 120\,s integrations (DDT
project 531; PI:\ Riechers). As the 6.22\,$\mu$m PAH feature in GN20
is redshifted to 31.44\,$\mu$m, we selected the first order of the
Long-Low module (LL1; 19.5--38.0\,$\mu$m) for this study. To improve
the calibrational accuracy, these ultra-deep observations were
obtained in spectral mapping mode, placing the target at six different
positions (separated by 20$''$) along the slit. High accuracy peak-ups
on isolated, bright Two Micron All Sky Survey (2MASS) stars were
obtained in the blue filter to optimize the pointing of the telescope.

The Basic Calibrated Data (BCD) files produced by the {\em Spitzer}
IRS S18.7.0 pipeline (including ramp fitting, dark sky subtraction,
droop and linearity corrections, flat-fielding, and wavelength
calibration) were used for further calibration and processing. The
two-dimensional dispersed frames were corrected for rogue pixels
(using IRSCLEAN), after which a small percentage of remaining bad
pixels was rejected. Then, the data were corrected for latent charge
buildup (dominantly from zodiacal background), which was fitted by a
constant polynomial and a time-variable component as a function of
wavelength and then subtracted. The sky background was removed by
fitting and subtracting a normalized `supersky' map (in which all
sources on the slit are masked) created from the data at different
slit positions within an astronomical observation request (AOR; see
P08 for details), resulting in minimal residual sky noise. After sky
removal, individual data frames were co-added using a clipped median,
and cleaned once more for persisting hot pixels. The Spitzer IRS
Custom Extraction (SPICE) program was then used to `blindly' extract
one-dimensional spectra from the two-dimensional coadds for each nod
position at the known 24\,$\mu$m position of the (unresolved) source
(which is too faint to be detected in individual two-dimensional
coadds). The uncertainty in the one-dimensional spectra was estimated
from extracting residual sky spectra in the off-source position of
each two-dimensional coadd.  The final spectrum was obtained by
averaging over all nod positions. The flux scale was obtained in SPICE
by extracting spectra of standard calibrator stars obtained during the
same campaign as our science observations, and found to be in good
agreement with existing 24\,$\mu$m photometry of the source. The final
noise on the spectrum between 21 and 32.5\,$\mu$m is estimated to be
40--90\,$\mu$Jy per 0.169\,$\mu$m wavelength bin. The spectrum was
trimmed above observed-frame wavelengths of $\sim$35\,$\mu$m due to
high noise. The nearby SMG GN20.2 was not covered by these
observations due to the IRS slit orientation at the time of
observation.

\section{Results}

\begin{figure*}[t]
\vspace{-4mm}

\epsscale{1.15}
\plotone{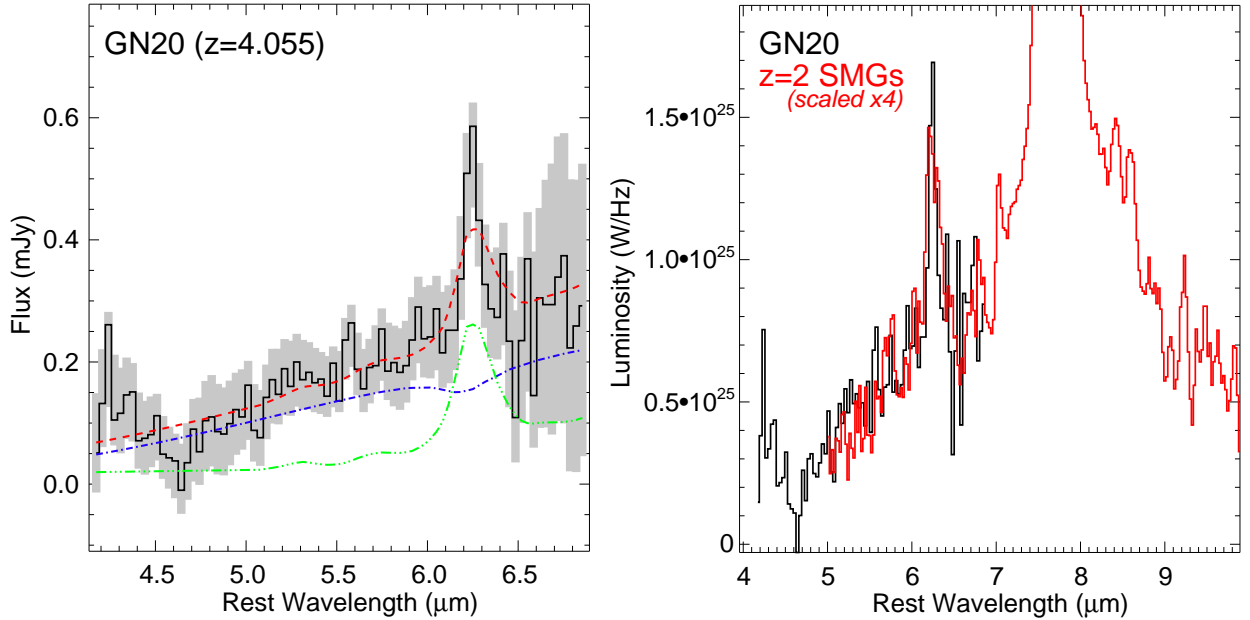}
\vspace{-3mm}

\caption{{\em Left}:\ {\em Spitzer} IRS spectrum of 6.2\,$\mu$m PAH
  and mid-IR continuum emission toward GN20 ($z$=4.055). The solid
  black histogram shows the raw (unsmoothed) spectrum, and the shaded
  region shows the associated $\pm$1$\sigma$ noise level (smoothed
  with a boxcar for display purposes), as estimated from the residual
  sky background. The dashed red line shows the best-fit SED, which is
  composed of a power-law component plus an extinction curve
  (dash-dotted blue line; the `dip' is due to extinction) and a
  starburst component (dash-triple dotted green line). At a rest-frame
  wavelength of 6\,$\mu$m, $\sim$2/3 (fit value:\ 66\%) of the
  continuum flux is due to the power-law component. {\em
    Right}:\ Comparison between GN20 (black histogram) and the
  $z$$\sim$2 SMG composite (which contains a $\sim$30\% contribution
  from a power-law AGN component) shown in Pope
  \etal\ (\citeyear{pop08}), scaled up by a factor of 4 in luminosity
  (red histogram).
\label{f1}}
%
\end{figure*}

We have detected 6.2\,$\mu$m PAH emission and 4--7\,$\mu$m continuum
emission toward the $z$=4.055 SMG GN20 (Fig.\ \ref{f1} {\em
  left}). The PAH emission extends over at least 5 wavelength bins,
and is detected at $>$6$\sigma$ significance (when simultaneously
fitting the underlying continuum). The continuum is detected at
typically 1.5$\sigma$--4$\sigma$ per wavelength bin over the full
spectral range ($>$80 bins), allowing us to constrain the continuum
slope over this spectral range.  The strength of the continuum
emission is consistent with the 24\,$\mu$m flux of
65.5$\pm$3.5\,$\mu$Jy measured with the Multiband Imaging Photometer
for Spitzer (MIPS; P08; D09) within the (considerable) relative
uncertainties.  When accounting for the steeply sloped continuum
emission, the PAH line shape is symmetric and consistent with a
Gaussian or Lorentzian profile, as expected.  A Gaussian fit to the
line profile and nearby continuum (i.e., the same method as used by
P08) yields an integrated 6.2\,$\mu$m PAH luminosity of $L_{\rm PAH
  (6.2\mu m)}$=(2.70$\pm$0.44)$\times$10$^{10}$\,\lsol, and a
rest-frame line equivalent width of 0.17$\pm$0.08\,$\mu$m.  A small
positive excess is seen close to the expected peak of the 5.7\,$\mu$m
PAH feature, but not enough to claim a detection. The spectrum also
shows a tentative decrement consistent with the wavelength of the
4.67\,$\mu$m $P$- and $R$-branches of the fundamental vibrational mode
of CO, which is observed in absorption toward some nearby
ultra-luminous infrared galaxies with embedded nuclei (e.g., Spoon
\etal\ \citeyear{spo04}, \citeyear{spo05}). Unfortunately, the
sensitivity is not sufficient to claim a detection of this CO
absorption feature.

We extract the remaining spectral properties of the source by fitting
a two-component model to the data, consisting of an extincted
power-law component (with a wavelength-dependent optical depth
obtained from the Draine \citeyear{dra03} extinction curves) and a
starburst component (see, e.g., Men\'endez-Delmestre
\etal\ \citeyear{men07}, \citeyear{men09}; P08). The starburst
component is a M82 template (F\"orster Schreiber
\etal\ \citeyear{fs03}), but adopting the starburst composite template
of Brandl \etal\ (\citeyear{bra06}) yields the same results within the
uncertainties.\footnote{The 6.2\,$\mu$m PAH to 6\,$\mu$m continuum
  flux ratio among the starbursts that are part of this template vary
  by about $\pm$20\%, which may be considered the minimum level of
  uncertainty in the decomposition. This is due to the fact that hot
  dust emission from a very compact, obscured starburst can also cause
  some level of excess in the mid-infrared continuum emission of a
  galaxy.} In comparison to the starburst template, the PAH feature is
both narrow and has a low equivalent width.\footnote{The 6.2\,$\mu$m
  PAH equivalent width for the starburst template reported by Brandl
  \etal\ (\citeyear{bra06}) is 0.53\,$\mu$m, with values in the range
  of 0.459 to 0.789\,$\mu$m for individual galaxies that are part of
  the template (measured using a method that gives similar results to
  our method, see discussion by P08).} Based on the examination of
subsets of the data, the narrow shape of the feature relative to
templates is of limited significance. Low PAH equivalent widths, such
as observed for GN20, are typically found in galaxies with significant
AGN emission at mid-infrared wavelengths (e.g., Armus
\etal\ \citeyear{arm07}).  Within the limited spectral range, the
spectral decomposition suggests that $\sim$2/3 (fit value:\ 66\%) of
the continuum emission at 6\,$\mu$m is due to the power-law component,
with only $\sim$1/3 (34\%) contribution from the starburst template,
with a level of uncertainty of at least 30\% from the fitting
alone. The power-law component alone, likely due to a hot dust source,
corresponds to a rest-frame 6\,$\mu$m continuum luminosity of
$\nu$$L_{\nu}$(6\,$\mu$m)=2.6$\times$10$^{45}$\,erg\,s$^{-1}$. As
discussed below, we interpret this power-law component to be
dominantly due to a heavily dust-enshrouded AGN in GN20, tracing the
thermally re-radiated emission from the obscuring material.

In the {\em right} panel of Fig.\ \ref{f1}, a comparison between the
spectrum of GN20 and the $z$$\sim$2 SMG composite spectrum of
P08\footnote{The 6.2\,$\mu$m PAH equivalent widths for individual SMGs
  that are part of this template are 0.22--1.10\,$\mu$m, which
  includes two upper limits of $<$0.45 and $<$0.50\,$\mu$m (P08).}
(scaled up by a factor of 4 in luminosity) is shown. Except for a
narrower PAH feature in GN20, both spectra look quite similar. The
average AGN contribution for the $z$$\sim$2 SMG composite sources is
$\sim$30\% (P08). This shows that the presence of an AGN in GN20 is
likely, but that, due to the limited spectral range of our
observations, the AGN contribution to the rest-frame 6\,$\mu$m
continuum luminosity is probably uncertain by about a factor of 2.

For comparison, we have extracted the rest-frame X-ray 2--10\,keV
luminosity of GN20 from the deep 2\,Ms exposure of the region obtained
with the {\em Chandra X-Ray Observatory} (Alexander
\etal\ \citeyear{ale03}). The source is not detected down to a
limiting luminosity of $L_{\rm
  X}$(2--10\,keV)$<$1.8$\times$10$^{42}$\,erg\,s$^{-1}$ (based on the
observed-frame 0.5--2\,keV flux limit, assuming a K-correction of
1.085 and a spectral slope of $\Gamma$=1.4). Under the assumption that
GN20 hosts a mid-infrared-luminous, obscured AGN, this X-ray limit
provides interesting constraints on its AGN properties (as outlined
below).

\section{Analysis}

\subsection{Spectral Energy Distribution}

\begin{figure}
\vspace{-3mm}

\epsscale{1.15}
\plotone{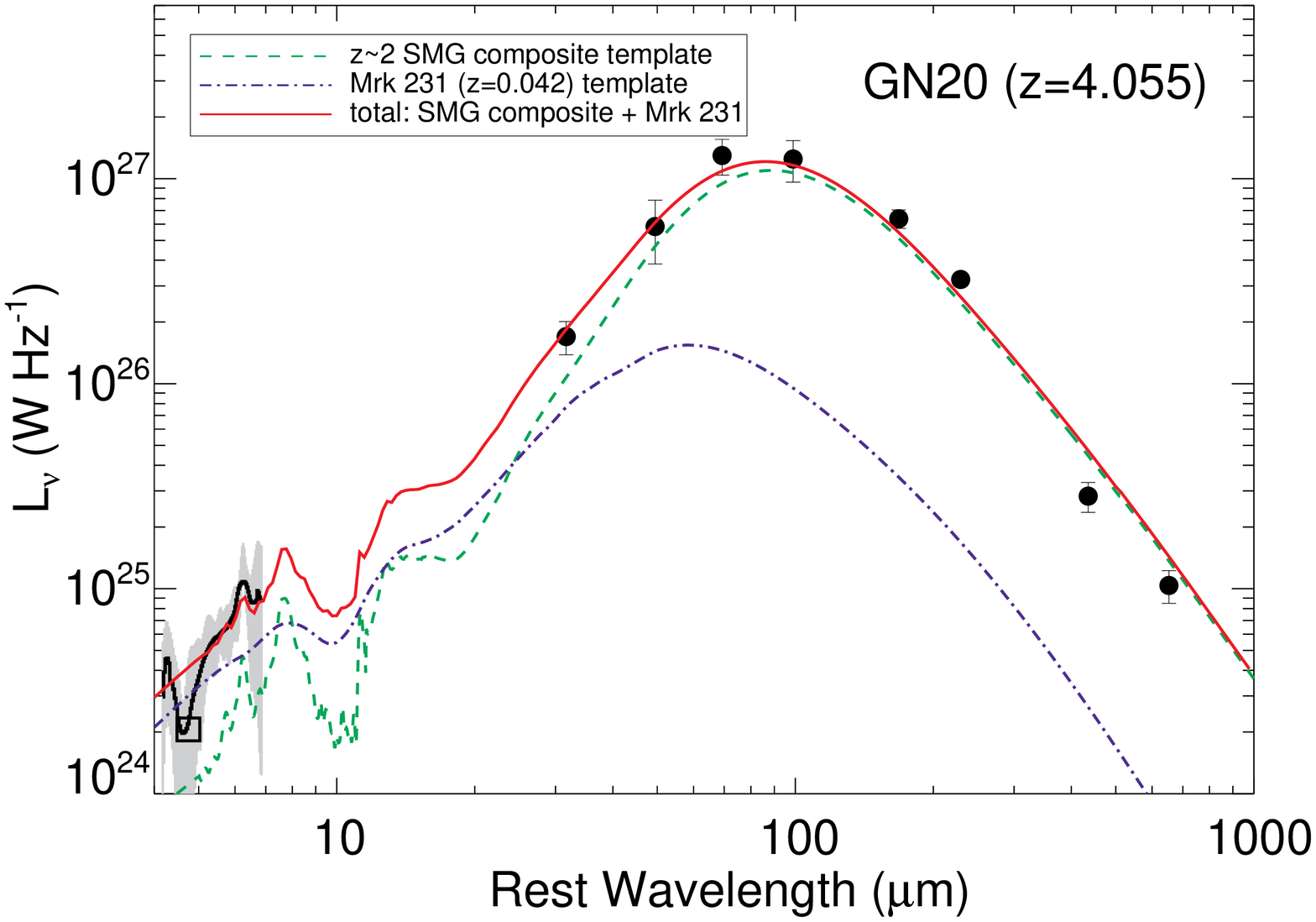}
\vspace{-3mm}

\caption{Spectral energy distribution of GN20 in specific luminosity
  ($L_{\nu}$), overlaid with the {\em Spitzer} IRS spectrum and
  spectral templates. The IRS spectrum (thick black line and gray
  shaded area) was smoothed for clarity. The {\em Spitzer} MIPS
  24\,$\mu$m photometry (65.5$\pm$3.5\,$\mu$Jy; open square; P08, D09)
  is shown for comparison. Rest-frame $>$30\,$\mu$m photometry (filled
  circles) is adopted from Magdis et al.\ (\citeyear{mag11}). The
  error bars of the {\em Herschel} SPIRE photometry are corrected for
  a contribution due to confusion noise (e.g., Nguyen et
  al.\ \citeyear{ngu10}). The green dashed line shows a smoothed
  composite SED template for $z$$\sim$2 SMGs (P08). The blue
  dash-dotted line shows an SED template of Mrk\,231 (Rigopoulou et
  al.\ \citeyear{rig99}; P08). The red long solid line shows a
  composite of both templates, fitted to the GN20 data by allowing a
  constant scaling factor for both components (i.e., SED$_{\rm
    total}$=$a_1$ SED$_{\rm SMG}$ + $a_2$ SED$_{\rm Mrk231}$).
\label{f2}}
%
\end{figure}

Our spectral decomposition suggests that $\sim$34\%--70\% of the
mid-infrared continuum flux in GN20 is due to star formation. This
suggests that, in order to determine the fraction of $L_{\rm IR}$ that
is due to star formation, the photometry at these wavelengths needs to
be corrected down accordingly to fit only the starburst fraction of
the spectral energy distribution (SED). We have applied this
correction to the SED fit presented by Magdis
\etal\ (\citeyear{mag11}; conservatively adopting the 34\% value from
the two-component fit).  This yields a starburst infrared luminosity
of $L_{\rm IR}$=(1.9$\pm$0.4)$\times$10$^{13}$\,\lsol, which is
essentially the same as the total infrared luminosity quoted by Magdis
\etal\ (\citeyear{mag11}). Given the SED shape of the source, the
impact of the correction (and its factor of $\sim$2 uncertainty) on
the total $L_{\rm IR}$ is minor. The correction does not account for a
potential cooler dust component that may be subject to AGN heating,
but that cannot be described by the mid-infrared power-law
component. At present, there is no evidence for the existence of such
an additional component in GN20.

To better quantify the possible contribution of any additional,
unconstrained component to $L_{\rm IR}$ if present, we simultaneously
fit the mid-infrared spectrum and full infrared SED of GN20 with a
composite full SED template of $z$$\sim$2 SMGs (consisting of the same
galaxies used in the mid-infrared composite in Fig.~\ref{f1}; P08) and
an SED template of the $z$=0.042 AGN-dominated infrared-luminous
galaxy Mrk\,231 (Fig.~\ref{f2}; Rigopoulou \etal\ \citeyear{rig99};
P08). The estimated contribution of star formation to the far-infrared
luminosity of Mrk\,231 is 30\%$\pm$15\% (Veilleux
\etal\ \citeyear{vei09}). This simultaneous composite fit reproduces
the overall SED properties of GN20 fairly well, and it suggests a
starburst contribution of 35\% to the mid-infrared luminosity. This
agrees well with the estimate based on the decomposition of the
mid-infrared spectrum alone. The fit overestimates the luminosity at
$\sim$4.4--5.0\,$\mu$m. This may be due to the possible presence of CO
absorption in this spectral region, but it could also indicate a lower
AGN contribution to the mid-infrared luminosity, consistent with the
range allowed by the mid-infrared spectral decomposition alone. The
fit also suggests that the AGN contribution to the total $L_{\rm IR}$
is $\lesssim$15\%--20\%. Given the underlying assumptions, we consider
this an approximate upper limit for the AGN contribution to $L_{\rm
  IR}$ in the following.

We can compare our results to those for individual $z$$\sim$2 SMGs. In
the sample of P08, three $z$$\sim$2 SMGs (classified as starbursts in
the mid-infrared) have upper limits on the AGN continuum contribution
to their mid-infrared spectra of $<$29\% to $<$35\%, and three other
$z$$\sim$2 SMGs (classified as AGN+starburst systems in the
mid-infrared) have estimates in the range of 47\%--61\%. For these
galaxies, which have mid-infrared AGN continuum fractions consistent
with GN20, P08 find upper limits of $<$15\% to $<$35\% and estimates
of 11\%--32\% for the AGN contribution to their total $L_{\rm IR}$ of
0.2--1.0$\times$10$^{13}$\,\lsol. For five $z$$\sim$2.5 SMGs selected
to have a steeply rising continuum slope between 4.5 and 8\,$\mu$m (as
expected for SMGs with a significant AGN contribution at those
wavelengths), Coppin \etal\ (\citeyear{cop10}) find 26\%--62\% AGN
contribution in the mid-infrared, and 23\%--36\% AGN contribution to
their total $L_{\rm IR}$ of
0.5--1.5$\times$10$^{13}$\,\lsol.\footnote{Systems in the samples of
  P08 and Coppin \etal\ (\citeyear{cop10}) with mid-infrared AGN
  continuum contributions that are not consistent with GN20 were
  excluded from this comparison.} This suggests a conservative limit
of $<$1/3 for a contribution of such an additional component to the
$L_{\rm IR}$ in GN20, if present. This is consistent with the results
from the composite SED fitting.

Based on the 1.4\,GHz radio flux density of 72$\pm$13\,$\mu$Jy (C10;
Pope \etal\ \citeyear{pop06}; Morrison \etal\ \citeyear{mor10}), we
find a $q$ parameter (i.e., ratio of $L_{\rm IR}$ to monochromatic
1.4\,GHz radio luminosity; Helou \etal\ \citeyear{hel85}) of $q$=2.39
for GN20 ($q$=2.32 if assuming an AGN contribution of 15\% to $L_{\rm
  IR}$), consistent with galaxies that follow the radio-infrared
correlation for nearby star-forming galaxies ($q$=2.3$\pm$0.1; Yun
\etal\ \citeyear{yun01}) and SMGs with solid radio identifications
($q$=2.4$\pm$0.1; Ivison \etal\ \citeyear{ivi10}). This is consistent
with no significant AGN contribution to the $L_{\rm IR}$ in GN20
beyond the corrections employed above. There also is no evidence for
an AGN contribution to the radio emission in GN20, in contrast to its
nearby companion galaxy, GN20.2a, which hosts a radio-detected AGN
(D09).

\subsection{AGN and Star Formation Relations}

\begin{figure}
\vspace{-1.75mm}

\epsscale{1.15}
\plotone{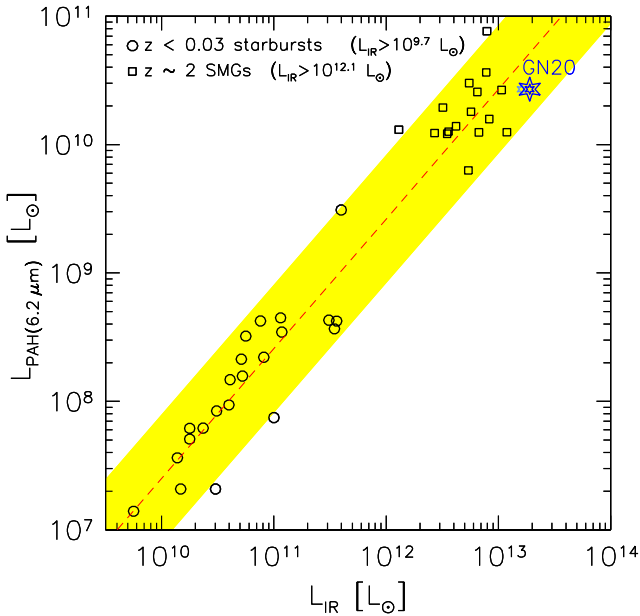}

\caption{Relation between $L_{\rm IR}$ (corrected for AGN
  contribution) and the 6.2\,$\mu$m PAH luminosity, $L_{\rm PAH
    (6.2\mu m)}$, for the local ($z$$<$0.03) starburst galaxy sample
  from Brandl \etal\ (\citeyear{bra06}; circles), the $z$$\sim$2 SMG
  sample from P08 and Coppin \etal\ (\citeyear{cop10}; squares), and
  the $z$=4.055 SMG GN20 (this work; stars). The line shows the
  best-fit relation found by P08:\ log\,$L_{\rm PAH (6.2\mu
    m)}$\ =\ 1.01 log\,$L_{\rm IR}$--2.7. The shaded region indicates
  a factor of 3 scatter. The small star shows the position of GN20
  when assuming a 15\% lower $L_{\rm IR}$.
\label{f3}}
%
\end{figure}

Interstellar PAH particles typically occur in photon-dominated regions
(PDRs), which are associated with star-forming clouds. In these
clouds, young stars heat large dust grains that emit re-processed
stellar light in the infrared.  Consequently, a relation between
$L_{\rm PAH (6.2\mu m)}$ and $L_{\rm IR}$ was found over a range in
luminosity from individual local star-forming regions up to starburst
galaxies (e.g., Peeters \etal\ \citeyear{pet04}) and even $z$$\sim$2
SMGs (P08). In Fig.~\ref{f3}, we plot this relation for nearby
starburst galaxies (Brandl \etal\ \citeyear{bra06}), the $z$$\sim$2
SMG sample from P08 and Coppin \etal\ (\citeyear{cop10}), and
GN20. All infrared luminosities are corrected for their estimated AGN
contribution. The (linear) relation shows a scatter by a factor of
$\sim$3 (as indicated by the shaded region), likely due to differences
in the physical properties of the individual star-forming environments
in these galaxies (such as starburst geometry and clumpiness, dust
distribution and temperature, or viewing angle toward a merger). The
$z$=4.055 SMG GN20 follows this relation within the scatter, providing
further evidence that its huge bolometric energy output is dominantly
due to an intense starburst. Given its high $L_{\rm IR}$ relative to
nearby starbursts, the relative strength of the PAH feature in GN20 is
consistent with a fairly large spatial extent of the starburst, but
relatively moderate dust temperatures (rather than a compact, hot
starburst region), such that typical PDRs giving rise to the PAH
emission are largely intact. This is in agreement with the moderate
molecular gas surface densities found for the bulk of the gas
reservoir (H12), and with the moderate characteristic dust temperature
found from SED fitting (33\,K; Magdis
\etal\ \citeyear{mag11}). Younger \etal\ (\citeyear{you08}) place a
limit of $\sim$4--8\,kpc (depending on the assumed source morphology)
on the FWHM diameter of the starburst in GN20, which is consistent
with a fairly large total extent as expected from the relative PAH
strength.

\begin{figure}
\epsscale{1.15}
\plotone{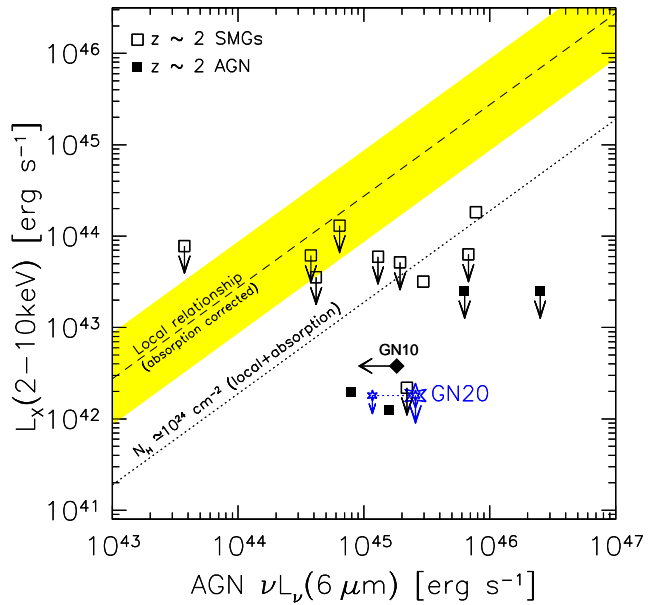}

\caption{Relation between the rest-frame mid-IR 6\,$\mu$m continuum
  luminosity (power-law component only) and the rest-frame X-ray
  2--10\,keV luminosity for the $z$$\sim$2 SMG sample from Coppin
  \etal\ (\citeyear{cop10}; empty squares) including C1 from P08, the
  $z$$\sim$2 AGN sample from Alexander \etal\ (\citeyear{ale08};
  filled squares), the $z$=4.042 SMG GN10 (Laird
  \etal\ \citeyear{lai10}; Pope \etal\ \citeyear{pop06}; an upper
  limit assuming 100\% power-law contribution to $L_{\nu}$(6\,$\mu$m)
  and the same mid-infrared continuum slope as GN20 is shown; filled
  diamond), and the $z$=4.055 SMG GN20 (this work; stars; the large
  symbol indicates the value from the spectral decomposition, the
  small symbol indicates an assumed power-law contribution of 30\% to
  $L_{\nu}$(6\,$\mu$m)). The black dashed line and shaded region
  indicate the absorption-corrected relation for nearby AGN (Lutz
  \etal\ \citeyear{lut04}). For comparison, the black dotted line
  indicates the expected luminosity ratio for an absorbing column
  density of $N_{\rm H}$$\simeq$10$^{24}$\,cm$^{-2}$, using the model
  by Alexander \etal\ (\citeyear{ale05}). Assuming that the local
  relation holds, sources significantly below this line (which include
  GN20) may be considered Compton-thick.
\label{f4}}
%
\end{figure}

In nearby AGN, hard X-ray emission from the central engine is commonly
considered to be a measure of the bolometric AGN luminosity, and thus,
should be linked to the strength of the thermally re-radiated
mid-infrared emission from hot dust close to the nucleus. This finding
has led to the discovery of an empirical relation between $L_{\rm
  X}$(2--10\,keV) and $\nu$$L_{\nu}$(6\,$\mu$m) (e.g., Lutz
\etal\ \citeyear{lut04}), which is linear over 3--5 orders of
magnitude, and extends out to high redshift. Galaxies can
significantly deviate from this relation due to intervening matter
along the line of sight to the AGN, i.e., due to a significant
absorbing column density of obscuring dust at X-ray wavelengths that
lowers the observed X-ray fluxes. In Fig.~\ref{f4}, the
(absorption-corrected) relation for nearby AGN is compared to a sample
of $z$$\sim$2 AGN and SMGs (Alexander \etal\ \citeyear{ale08}; Coppin
\etal\ \citeyear{cop10}), the $z$=4.042 SMG GN10 (Laird
\etal\ \citeyear{lai10}; Pope \etal\ \citeyear{pop06}), and GN20. For
all galaxies, only the estimated AGN power-law contribution to the
mid-infrared emission from the spectral decomposition is shown (for
GN10, only an upper limit exists due to the lack of mid-infrared
spectroscopy). The deep X-ray limit indicates that GN20 would be
obscured by a factor of $>$430 if the local relation were to hold
(with a factor of at least $\sim$3 uncertainty due to the scatter in
that relation, as indicated by the shaded region in Fig.~\ref{f4}, and
some additional uncertainty in the K-correction and X-ray spectral
slope assumed for GN20). This suggests that, if an AGN contributes at
$\sim$30\%--66\% level to the mid-infrared continuum emission (as
indicated by the spectral decomposition), it likely is a type-2 AGN
obscured by Compton-thick material.  This would be in agreement with
its non-detection in the optical, and with the extreme obscuration of
the massive, gas- and dust-rich host galaxy (e.g., D09; C10;
H12). This AGN is unlikely to be very luminous:\ even when assuming a
correction factor of $\gtrsim$430, the limit on $L_{\rm
  X}$(2--10\,keV) corresponds to only $\sim$1\% of $L_{\rm IR}$, which
is far below the values typically found for luminous AGN (Alexander
\etal\ \citeyear{ale05}).

\subsection{Star Formation Rate}

Based on the relation between $L_{\rm PAH (6.2\mu m)}$ and $L_{\rm
  IR}$, P08 have determined a relation to convert $L_{\rm PAH (6.2\mu
  m)}$ into a star formation rate (SFR; their Eq.~7). For GN20, this
relation suggests a SFR of $\sim$1600\,\msol\,yr$^{-1}$ when assuming
a Chabrier (\citeyear{cha03}) stellar initial mass function
(IMF). With the same IMF, the $L_{\rm IR}$ of GN20 suggests a SFR of
1900\,\msol\,yr$^{-1}$ ($\sim$1600\,\msol\,yr$^{-1}$ if assuming a
15\% AGN contribution to $L_{\rm IR}$), comparable to that of the
$z$=5.3 SMG AzTEC-3 (which also resides in a protocluster environment;
Capak \etal\ \citeyear{cap11}; Riechers
\etal\ \citeyear{rie10}). These estimates thus consistently suggest a
SFR of at least $\sim$1600\,\msol\,yr$^{-1}$ within the relative
uncertainties.

Given the large molecular gas mass of
1.8$\times$10$^{11}$\,($\alpha_{\rm CO}$/1.1)\,\msol\ (H12), this
corresponds to a gas consumption timescale of $\sim$10$^8$\,yr. This
is consistent with a short, intense burst of star formation, as seen
in `typical' $z$$\sim$2 SMGs (e.g., Greve \etal\ \citeyear{gre05}) and
$z$$\gtrsim$4 quasar host galaxies (e.g., Riechers
\etal\ \citeyear{rie08a}, \citeyear{rie08b}).

\section{Discussion}

We have detected 6.2\,$\mu$m PAH and rest-frame 4--7\,$\mu$m continuum
emission toward the $z$=4.055 SMG GN20, using the {\em Spitzer}
IRS. GN20 is one of the most luminous dusty galaxies known at high
redshift. The 6.2\,$\mu$m PAH equivalent width is a few times lower
than in `typical' $z$$\sim$2 SMGs, due to a relatively high
contribution ($\sim$1/3--2/3) of a power-law component to the
continuum emission.  When fit in the same fashion, $z$$\sim$2 SMGs
typically show $<$30\% contribution of power-law continuum emission in
the mid-infrared ($\sim$80\% of the population; e.g., P08).

The simplest compelling interpretation is that the different
characteristics of GN20 relative to nearby starbursts and the majority
of `typical' $z$$\sim$2 SMGs are related to a luminous power source
besides the starburst, with a possible contribution from other
mechanisms and/or due to unusual properties of the dust. This
additional, luminous power source is likely an obscured AGN. Given the
lack of direct evidence for an AGN at other wavelengths, this
interpretation warrants further exploration.

On the one hand, there is no known mechanism other than AGN activity
that can produce the high excess in mid-infrared continuum luminosity
observed in GN20 without making fairly extreme assumptions.  The
tentatively identified 4.67\,$\mu$m CO absorption feature, if
confirmed, would be consistent with a very warm, embedded source
contributing significantly to the 6\,$\mu$m continuum emission
(against which the absorption is measured), such as an obscured
AGN. In principle, a very compact, obscured starburst can cause some
excess in the mid-infrared continuum emission of a galaxy, which in
fact may be responsible for at least part of the variations seen among
the mid-infrared spectra of nearby starbursts. However, GN20 is
dominated by a moderately warm, spatially extended starburst, without
evidence for a significant compact, hot subcomponent. It has been
argued in the literature that small, hot dust grains in the most
energetic H{\scriptsize II} regions can produce power-law-shaped
radiation in the mid-infrared (e.g., Genzel \etal\ \citeyear{gen98};
Tran \etal\ \citeyear{tra01}). Due to the high redshift of GN20 and
the likely presence of multiple emission components, the spectral
coverage of the IRS spectrum is not sufficient to firmly distinguish
between the flatter mid-infrared power-law slopes of emission
associated with a dust torus surrounding an AGN, and the steeper
slopes observed towards H{\scriptsize II} regions with highly
concentrated star formation. Given the overall cool dust SED shape of
GN20, it remains unclear if such H{\scriptsize II} regions are
sufficiently common to produce a power-law-shaped excess component as
strong as observed.  In any case, given a possible contribution from
H{\scriptsize II} regions, we consider our estimates for the
power-law-shaped component to be an upper limit for the actual AGN
contribution.

On the other hand, there are several potential reasons for the low
observed equivalent width of the PAH feature. One possibility to
interpret the low equivalent width of the PAH feature is that the
adopted nearby starburst template (or even the $z$$\sim$2 SMG
template) may not be suitable to fit the spectrum of substantially
more luminous and/or distant starburst galaxies at $z$$>$4. Besides
the relative strength of the hot dust continuum emission, the
equivalent width of the 6.2\,$\,\mu$m PAH feature depends on the
ionized vs.\ neutral PAH ratio, which in turn depends on metallicity
and the hardness of the radiation field that the PAH particles are
bathed in (e.g., Draine \& Li \citeyear{dl07}). The relative strength
of the CO emission in GN20 compared to its far-infrared dust
properties (D09; C10; H12), however, does not indicate substantial
differences compared to $z$$\sim$2 SMGs (e.g., Tacconi
\etal\ \citeyear{tac08}; Riechers et al.\ \citeyear{rie11a},
\citeyear{rie11b}) in terms of dust/gas-phase metallicity.
Furthermore, the overall radiation field that penetrates the dust
would be substantially harder if composed of both a starburst and an
AGN component. This could be consistent with a non-standard ionized
vs.\ neutral PAH ratio, which may result in a narrower feature, and a
lower equivalent width, as seen in GN20 (it however may be difficult
to reconcile a scenario where the AGN radiation penetrates the opaque,
obscuring dust and gas throughout the large, 14\,kpc diameter of the
reservoir).

Interestingly, the $z$$\sim$2 SMG with the lowest equivalent width
6.2\,$\mu$m PAH feature (C1, 0.05$\pm$0.01\,$\mu$m; i.e., $<$1/3 the
equivalent width of GN20) found by P08, like GN20, is consistent with
hosting a Compton-thick AGN. Studies of other types of galaxies find
that systems with 6.2\,$\mu$m PAH equivalent widths of $<$0.2\,$\mu$m
are typically AGN-dominated in the mid-infrared (e.g., Armus
\etal\ \citeyear{arm07}; Sajina \etal\ \citeyear{saj07}). Also, Coppin
\etal\ (\citeyear{cop10}) have studied a sample of $z$$\sim$2 SMGs
that were suspected to be dominated by AGN (representing $<$15\% of
the $z$$\sim$2 SMG population),\footnote{With $S_{\rm 8\mu m}$/$S_{\rm
    4.5\mu m}$=2.75, GN20 formally follows the color excess (caused by
  500--1000\,K hot dust) selection criterion for mid-infrared
  AGN-dominated SMGs as defined by Coppin et al., but only due to the
  shorter rest-frame wavelengths probed in these bands at $z$$>$4
  relative to their $z$$\sim$2.5 sample.} several of which are at
least consistent with being Compton-thick AGN. These studies find that
in a number of these systems (corresponding to $\lesssim$5\% of the
SMG population, not including the $\sim$4\% of SMGs that are type-1
AGN/quasars), the AGN is bolometrically important, and sometimes even
dominant, relative to their submillimeter-detected luminous starbursts
of few hundreds to $>$1000\,\msol\,yr$^{-1}$.  Our findings show that
an obscured, Compton-thick AGN may significantly contribute to, and
perhaps dominate the energy output in the mid-infrared in GN20, but it
appears to only yield a minor contribution to the bolometric energy
output of the galaxy.

In conclusion, we have used the {\em Spitzer} IRS to constrain the AGN
and star formation properties in a starburst galaxy at $z$$>$4,
employing the unique diagnostics observable only at mid-infrared
wavelengths.  Based on the detection of 6.2\,$\mu$m PAH and
4--7\,$\mu$m continuum emission in the exceptional $z$=4.055 SMG GN20,
we have found evidence for the likely presence of a heavily buried,
Compton-thick, but bolometrically not dominant AGN hosted by a
massive, dust-obscured $\sim$1600--1900\,\msol\,yr$^{-1}$ starburst.

\acknowledgments

We thank Dave Alexander for help with obtaining the X-ray limit, and
for communicating some data source tables, Henrik Spoon for helpful
comments, Georgios Magdis for help with an earlier version of
Figure~\ref{f2}, and the anonymous referee for a helpful and
constructive report.  DR acknowledges support from a {\em Spitzer
  Space Telescope} grant related to this project.  DR appreciates the
hospitality at the Aspen Center for Physics, where part of this
manuscript was written.  This work is based on observations made with
the {\em Spitzer Space Telescope}, which is operated by the Jet
Propulsion Laboratory, California Institute of Technology, under a
contract with NASA. The IRS was a collaborative venture between
Cornell University and Ball Aerospace Corporation funded by NASA
through the Jet Propulsion Laboratory and Ames Research Center.  The
scientific results reported in this article are based in part on
observations made by the {\em Chandra X-ray Observatory}.

\appendix

\subsection{Black Hole Mass and Accretion Rate}

Assuming that GN20 hosts an obscured AGN, and that the local
$\nu$$L_{\nu}$(6\,$\mu$m)--$L_{\rm X}$(2--10\,keV) relation holds,
i.e., that the large absorption corrections found in Sect.~4.2 are
valid, we can attempt to derive an Eddington limit for the black hole
mass $M_{\rm BH}$ of GN20 based on $\nu$$L_{\nu}$(6\,$\mu$m). Assuming
a bolometric correction of $\kappa_{\rm 2-10keV}$=$L_{\rm
  bol}$/$L_{\rm X}$(2--10\,keV)=55 (for high $L_{\rm bol}$/$L_{\rm
  edd}$ galaxies; Vasudevan \& Fabian \citeyear{vf07}), we find
$M_{\rm BH}^{\rm Edd}$$\simeq$1.4--3.0$\times$10$^8$\,\msol\ for
30\%--66\% of $\nu$$L_{\nu}$(6\,$\mu$m), corresponding to
$\sim$0.03\%--0.06\% of the total (dynamical) mass of the galaxy
($M_{\rm dyn}$=5.4$\pm$2.4$\times$10$^{11}$\,\msol; H12).  This ratio
is comparable to what is found in $z$$\sim$2 SMGs ($\sim$0.05\%;
Alexander \etal\ \citeyear{ale05b}, \citeyear{ale08}).  This may imply
that the $M_{\rm BH}$ of GN20 falls short of the near-linear local
$M_{\rm BH}$--$M_{\rm bulge}$ relation (which corresponds to a mass
ratio of 0.19\% at the $M_{\rm dyn}$ of GN20; Magorrian
\etal\ \citeyear{mag98}; H\"aring \& Rix \citeyear{hr04}) by a factor
of a few, but this offset is of the same order of magnitude as the
uncertainties. In any case, these considerations show that the
supermassive black hole properties implied by the above estimates
appear to be reasonable. Assuming a canonical mass accretion
efficiency of $\eta$=0.1, this also corresponds to an accretion rate
of $\dot{M}$$\simeq$4--7\,\msol\,yr$^{-1}$, or $\sim$0.2\%--0.4\% of
the SFR.  Given the underlying corrections and uncertainties, we
consider these (plausible) values to be reliable to within a factor of
3--5 at best.\\


\begin{thebibliography}{}

\bibitem[Alexander et al.(2003)]{ale03} Alexander, D.~M., et al.\ 2003, AJ, 126, 539
\bibitem[Alexander et al.(2005a)]{ale05} Alexander, D.~M., et al.\ 2005a, ApJ, 632, 736
\bibitem[Alexander et al.(2005b)]{ale05b} Alexander, D.~M., et al.\ 2005b, Nature, 434, 738
\bibitem[Alexander et al.(2008)]{ale08} Alexander, D.~M., et al.\ 2008, AJ, 135, 1968
\bibitem[Armus et al.(2007)]{arm07} Armus, L., et al.\ 2007, ApJ, 656, 148
\bibitem[Blain et al.(2002)]{bla02} Blain, A.~W., et al.\ 2002, PhR, 369, 111
\bibitem[Brandl et al.(2006)]{bra06} Brandl, B.~R., et al.\ 2006, ApJ, 653, 1129
\bibitem[Capak et al.(2008)]{cap08} Capak, P., et al.\ 2008, ApJ, 681, L53
\bibitem[Capak et al.(2011)]{cap11} Capak, P., et al.\ 2011, Nature, 470, 233
\bibitem[Carilli et al.(2010)]{car10} Carilli, C.~L., Daddi, E., Riechers, D., et al.\ 2010, ApJ, 714, 1407 [C10]
\bibitem[Chabrier(2003)]{cha03} Chabrier, G.\ 2003, PASP, 115, 763
\bibitem[Coppin et al.(2009)]{cop09} Coppin, K., et al.\ 2009, MNRAS, 395, 1905
\bibitem[Coppin et al.(2010)]{cop10} Coppin, K., et al.\ 2010, ApJ, 713, 503
\bibitem[Daddi et al.(2007)]{dad07} Daddi, E., et al.\ 2007, ApJ, 670, 156
\bibitem[Daddi et al.(2009a)]{dad09} Daddi, E., et al.\ 2009a, ApJ, 694, 1517 [D09]
\bibitem[Daddi et al.(2009b)]{dad09b} Daddi, E., et al.\ 2009b, ApJ, 695, L176
\bibitem[Draine(2003)]{dra03} Draine, B.~T.\ 2003, ARA\&A, 41, 241
\bibitem[Draine\&Li(2007)]{dl07} Draine, B.~T., \& Li, A.\ 2007, ApJ, 657, 810
\bibitem[Farrah et al.(2007)]{far07} Farrah, D., et al.\ 2007, ApJ, 667, 149
\bibitem[F\"orster Schreiber et al.(2003)]{fs03} F\"orster Schreiber, N.~M., et al.\ 2003, A\&A, 399, 833
\bibitem[Genzel et al.(1998)]{gen98} Genzel, R., et al.\ 1998, ApJ, 498, 579
\bibitem[Greve et al.(2005)]{gre05} Greve, T.~R., et al.\ 2005, MNRAS, 359, 1165
\bibitem[H\"aring \& Rix(2004)]{hr04} H\"aring, N., \& Rix, H.-W.\ 2004, ApJ, 604, L89
\bibitem[Helou et al.(1985)]{hel85} Helou, G., et al.\ 1985, ApJ, 298, L7
\bibitem[Hodge et al.(2012)]{hod12} Hodge, J., et al.\ 2012, ApJ, 760, 11 [H12]
\bibitem[Hodge et al.(2013)]{hod13} Hodge, J., Carilli, C.~L., Walter, F., Daddi, E., \& Riechers, D.\ 2013, ApJ, 776, 22
\bibitem[Houck et al.(2004)]{hou04} Houck, J.~R., et al.\ 2004, ApJS, 154, 18
\bibitem[Ivison et al.(2010)]{ivi10} Ivison, R.~J., et al.\ 2010, A\&A, 518, L31
\bibitem[Ivison et al.(2011)]{ivi11} Ivison, R.~J., et al.\ 2011, MNRAS, 412, 1913
\bibitem[Iono et al.(2006)]{ion06} Iono, D., et al.\ 2006, ApJ, 640, L1
\bibitem[Laird et al.(2010)]{lai10} Laird, E.~S., Nandra, K., Pope, A., \& Scott, D.\ 2010, MNRAS, 401, 2763
\bibitem[Lutz et al.(2004)]{lut04} Lutz, D., et al.\ 2004, A\&A, 418, 465
\bibitem[Magdis et al.(2011)]{mag11} Magdis, G.~E., et al. 2011, ApJ, 740, L15
\bibitem[Magorrian et al.(1998)]{mag98} Magorrian, J., et al.\ 1998, AJ, 115, 2285
\bibitem[Menendez-Delmestre et al.(2007)]{men07} Men\'endez-Delmestre, K., et al.\ 2007, ApJ, 655, L65
\bibitem[Menendez-Delmestre et al.(2009)]{men09} Men\'endez-Delmestre, K., et al.\ 2009, ApJ, 699, 667
\bibitem[Morrison et al.(2010)]{mor10} Morrison, G.~E., Owen, F.~N., Dickinson, M., Ivison, R.~J., \& Ibar, E.\ 2010, ApJS, 188, 178
\bibitem[Nguyen et al.(2010)]{ngu10} Nguyen, H.~T., et al.\ 2010, A\&A, 518, L5
\bibitem[Peeters et al.(2004)]{pet04} Peeters, E., et al.\ 2004, ApJ, 613, 986
\bibitem[Petric et al.(2011)]{pet11} Petric, A., et al.\ 2011, ApJ, 730, 28
\bibitem[Pope et al.(2005)]{pop05} Pope, A., et al.\ 2005, MNRAS, 358, 149
\bibitem[Pope et al.(2006)]{pop06} Pope, A., et al.\ 2006, MNRAS, 370, 1185
\bibitem[Pope et al.(2008)]{pop08} Pope, A., et al.\ 2008, ApJ, 675, 1171 [P08]
\bibitem[Renzini(2006)]{ren06} Renzini, A.\ 2006, ARA\&A, 44, 141
\bibitem[Riechers et al.(2008a)]{rie08a} Riechers, D.~A., et al.\ 2008a, ApJ, 686, L9
\bibitem[Riechers et al.(2008b)]{rie08b} Riechers, D.~A., et al.\ 2008b, ApJ, 686, 851
\bibitem[Riechers et al.(2010)]{rie10} Riechers, D.~A., et al.\ 2010, ApJ, 720, L131
\bibitem[Riechers et al.(2011a)]{rie11a} Riechers, D.~A., et al.\ 2011a, ApJ, 733, L11
\bibitem[Riechers et al.(2011b)]{rie11b} Riechers, D.~A., Hodge, J., Walter, F., Carilli, C.~L., \& Bertoldi, F.\ 2011b, ApJ, 739, L31
\bibitem[Riechers et al.(2013)]{rie13} Riechers, D.~A., et al.\ 2013, Nature, 496, 329
\bibitem[Rigopoulou et al.(1999)]{rig99} Rigopoulou, D., Spoon, H.~W.~W., Genzel, R., Lutz, D., Moorwood, A.~F.~M., \& Tran, Q.~D.\ 1999, AJ, 118, 2625
\bibitem[Sajina et al.(2007)]{saj07} Sajina, A., et al.\ 2007, ApJ, 664, 713
\bibitem[Spergel et al.(2003)]{spe03} Spergel, D.~N., Verde, L., Peiris, H.~V., \etal\ 2003, ApJS, 148, 175
\bibitem[Spergel et al.(2007)]{spe07} Spergel, D.~N., Bean, R., Dor\'e, O., \etal\ 2007, ApJS, 170, 377
\bibitem[Spoon et al.(2004)]{spo04} Spoon, H.~W.~W., et al.\ 2004, ApJS, 154, 184
\bibitem[Spoon et al.(2005)]{spo05} Spoon, H.~W.~W., et al.\ 2005, IAUS, 231, 281
\bibitem[Taccconi et al.(2008)]{tac08} Tacconi, L.~J., et al.\ 2008, ApJ, 680, 246
\bibitem[Tran et al.(2001)]{tra01} Tran, Q.~D., et al.\ 2001, ApJ, 552, 527
\bibitem[Vasudevan \& Fabian(2007)]{vf07} Vasudevan, R.~V., \& Fabian, A.~C.\ 2007, MNRAS, 381, 1235
\bibitem[Veilleux et al.(2009)]{vei09} Veilleux, S., et al.\ 2009, ApJS, 182, 628
\bibitem[Younger et al.(2008)]{you08} Younger, J.~D., et al.\ 2008, ApJ, 688, 59
\bibitem[Yun et al.(2001)]{yun01} Yun, M.~S., Reddy, N.~A., \& Condon, J.~J.\ 2001, ApJ, 554, 803
\end{thebibliography}
\end{document}